\documentstyle[prb,aps,eqsecnum,multicol,ifthen,epsf]{revtex}
\tighten

\newcommand{\wide}[2]{                                                        %
\end{multicols}                                                               %
\widetext                                                                     %
\noindent                                                                     %
\ifthenelse{\equal{#1}{t}}                                                    %
{}                                                                            %
{                                                                             %
\raisebox{0.1in}[0in][0.02in]{$\rule{3.575in}{0.002in}                        %
\rule{0.002in}{0.08in}$}                                                      %
}                                                                             %
#2                                                                            %
\ifthenelse{\equal{#1}{b}}                                                    %
{}                                                                            %
{                                                                             %
{\raisebox{-0.1in}[0in][0.02in]                                               %
{\hspace{3.575in}$\rule{0.002in}{0.08in}                                      %
\rule[0.08in]{3.575in}{0.002in}$}                                             %
}                                                                             %
}                                                                             %
\begin{multicols}{2}                                                          %
\noindent                                                                     %
}                                                                             %
\begin{document}

\draft
\preprint{submitted to: J. Magn. Magn. Mater.}

\title{Influence of disorder on the perpendicular 
magnetoresistance of magnetic multilayers}

\author{P.~Bruno}

\address{Max-Planck-Institut f\"ur Mikrostrukturphysik, Weinberg 2, D-06120 
Halle, Germany}

\author{H.~Itoh, J.~Inoue, and S.~Nonoyama}

\address{Department of Applied Physics, Nagoya University, Nagoya 464-01,
Japan}

\date{14 August 1998}

\maketitle

\begin{abstract}
The effect of disorder on the perpendicular magnetoresistance of magnetic 
multilayers is investigated theoretically. Various 
kinds of disorder are considered: (i) interface substitutional disorder and 
(ii) bulk disorder in the various layers and in the leads. The calculations are 
based upon the non-equilibrium Green's function formalism, together with the 
recursion method for calculating the real-space Green's function.
\end{abstract}

\pacs{\normalsize published in: J. Magn. Magn. Mater. {\bf 198-199}, 
46--48 (1999).}

\begin{multicols}{2}

\section{Introduction}

The common wisdom about giant magnetoresistance (GMR) is that it is due to spin
dependent scattering.\cite{ Baibich1988} Although the transport is
certainly diffusive for most systems investigated experimentally, it has been
pointed out that pure ballistic transport as well would lead to GMR for the
perpendicular current geometry.\cite{ Schep1995} Subsequently, it was proposed
that for ballistic perpendicular transport, the electron confinement within the
layers would yield a quantum size effect, i.e., an oscillatory behavior of the
conductance and of the GMR as a function of layer thickness.\cite{ Barnas1995,
Mathon1995} 

In the present paper, we investigate the influence of disorder on perpendicular
magneto-resistance, and discuss the crossover from ballistic to diffusive 
transport as disorder increases.

\section{Model}

The model considered here consists of two ferromagnetic layers of 
thickness $L$ separated by a non-magnetic spacer layer of thickness $D$; the 
structure is sandwiched between two ideal leads connected to ideal reservoirs. 
We use a simple-cubic single-band tight-binding Hamiltonian with a 
nearest-neighbor
hopping. The hopping parameter, $t$, is the same for all pairs of nearest 
neighbors 
(for numerical calculations, $t=-1$ will be assumed). The on-site energy of 
site $i$,
$\varepsilon_i$, depends on the
chemical nature of the site and (for magnetic sites) on its spin. The majority 
spin on-site 
energy of the magnetic sites is taken equal to the on-site energy of 
nonmagnetic sites,
and the minority spin on-site energy of the magnetic sites is 0.3~$|t|$ higher 
than 
the latter. The Fermi level is taken 0.8~$|t|$ above the 
bottom of the band; this corresponds to a quasi-free-electron case, with 
quasi-parabolic 
band structure and quasi-spherical Fermi surface.

Two kinds of models for disorder are considered: (i) random substitutional 
disorder, 
and (ii) the Anderson model of disorder. Model (i) is particularly well suited 
to 
study the effect of interface interdiffusion. In model (ii), a random on-site 
energy
(characterized by a square distribution of width $W$) is added to the on-site 
energy
of the perfect case; this is a convenient model for bulk disorder. For both 
cases, 
a periodic (along the in-plane directions) model of disorder with $10\times 10$ 
supercell is considered. We did not perform systematically an average over 
disorder
configurations, but we checked that this 100-sites supercell is large enough for 
fluctuations to be unimportant. For ${\bf k}_\|$ summations, we take 100 
${\bf k}_\|$-points in the two-dimensional Brillouin zone corresponding to  the 
supercell;
this sampling is equivalent to taking $10^4$ ${\bf k}_\|$-points for the 
perfect case.

\section{Theory}

The calculation of the conductance of the structure is based upon the 
non-equilibrium 
Green's function formalism.\cite{ Kadanoff1962} When applied to the case of a 
two-terminal 
ballistic mesoscopic conductor 
we obtain the following result for the conductance:\cite{ Datta1990, Bruno1998} 
\begin{equation}\label{ eq:conductance}
G = \frac{e^2}{h} \mbox{Tr}_{{\bf k}_\| ,\sigma} \left( \Gamma_l\, G^R_{lr}\, 
\Gamma_r \, G^A_{rl} \right)
\end{equation}
In the above equation, the upper indices $R$ and $A$ refer to the retarded and 
advanced Green's function, respectively; the lower $l$ and $r$ indices refer to 
the left
and right reservoirs, respectively. Thus, $G_{lr}^R$ is the off-diagonal 
retarded Green's 
function linking the left reservoir to the right reservoir; $\Gamma_l$ 
(respectively, 
$\Gamma_r$) are given by $\Gamma_{l(r)} = t^2 \, A_{l(r)}$,
where $A_l$ (respectively, $A_r$) is the spectral density in the first plane of 
the 
left (respectively, right) reservoir, when it is decoupled from the conductor. 
All the
quantities in Eq.~(\ref{ eq:conductance}) are taken at the Fermi level. The 
conductance obtained here from the non-equilibrium Green's function formalism, 
Eq.~(\ref{ eq:conductance}), can be shown to be equivalent to the ones 
obtained from 
the Kubo formula, and from the Landauer-B\"uttiker formalism.\cite{ Bruno1998} 

By separating the ${\bf k}_\|$-conserving and ${\bf k}_\|$-non-conserving terms, 
we can separate the ballistic and diffusive contributions to the total 
conductance: 
\wide{t}{
\begin{mathletters}
\begin{eqnarray}
G_{\rm bal} &\equiv & \frac{e^2}{h} \sum_{{\bf k}_\|} \mbox{Tr}_\sigma 
\left[ \Gamma_l({\bf k}_\| )\, G_{lr}^R ({\bf k}_\| , {\bf k}_\| ) \,
       \Gamma_r({\bf k}_\| ) \, G_{rl}^A ({\bf k}_\| , {\bf k}_\| ) \right] ,
\\
G_{\rm dif} &\equiv & \frac{e^2}{h} \sum_{{\bf k}_\| \neq {\bf k}_\|^\prime } 
\mbox{Tr}_\sigma 
\left[ \Gamma_l({\bf k}_\| )\, G_{lr}^R ({\bf k}_\| , {\bf k}_\|^\prime ) \,
       \Gamma_r({\bf k}_\|^\prime ) \, G_{rl}^A ({\bf k}_\|^\prime ,{\bf k}_\| ) 
\right] .
\end{eqnarray}
\end{mathletters}
}

\narrowtext{
\begin{figure}[h]
%\vspace*{2cm}
\epsfxsize=8cm
\epsffile{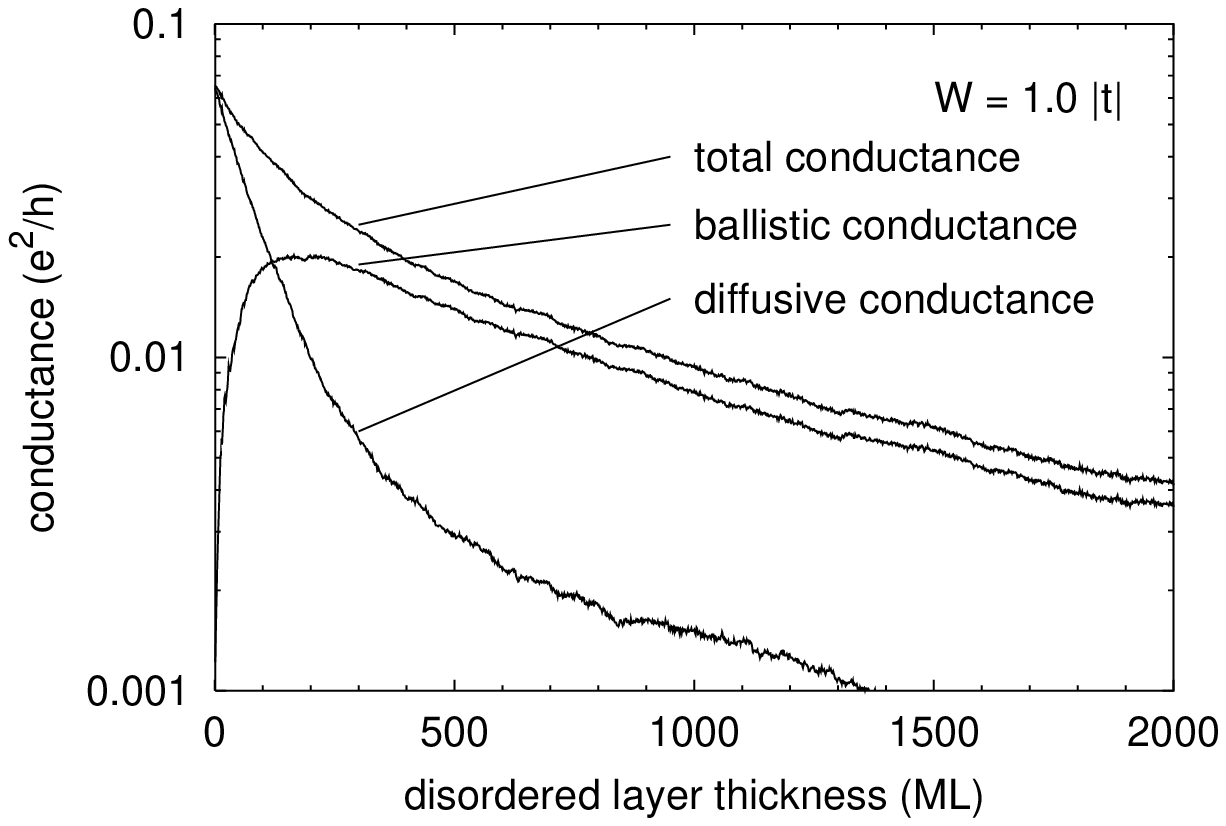}
%\end{center}
%\vspace*{\baselineskip}
\caption{Conductance of a disordered nonmagnetic layer as a function of layer 
thickness.}
\label{ fig:G}
\end{figure}
}

\section{Results and discussion}

Before considering the problem of perpendicular magnetoresistance, we illustrate
our method by presenting the conductance of a nonmagnetic layer of thickness up 
to
2000 monolayers (ML), with an Anderson-like disorder of width $W=1.0 |t|$. The 
results 
are shown
in Fig.~\ref{ fig:G}. For small thicknesses, the transport is quasi-ballistic, 
but for
thicknesses larger than the mean-free path (approximately 400~ML here), the 
transport 
becomes diffusive.

Let us now consider the problem of the perpendicular magneto-resistance. We 
define the 
magnetoresistance as\cite{ note} $A \equiv (G_F - G_{AF})/(G_F + G_{AF})$.
The results are displayed in Fig.~\ref{ fig:MR}. 
The magneto-resistance of the perfect system (Fig.~\ref{ fig:MR}a) exhibits 
the oscillations due 
to the 
quantum size effect.\cite{ Mathon1995} 
We have investigated the case of interface interdiffusion (Fig.~\ref{ fig:MR}b) 
by including two 
planes of random substitutional alloy: it appears that the latter influences 
only weakly the magnetoresistance. We have also performed calculations treating 
the interface alloy within the virtual crystal approximation: the 
magnetoresistance 
(not shown here) is almost identical to the ``exact'' one (Fig.~\ref{ fig:MR}b);
the effect of is interface interdiffusion is thus merely to replace the abrupt 
interface by a graded interface, which reduces the reflectivity, but produces 
little diffuse scattering.

If we include bulk disorder in the spacer (Anderson model of 
disorder) the diffuse scattering leads to 
reduction
of the magneto-resistance for large spacer thickness, but the quantum size 
effect is 
still present (Fig.~\ref{ fig:MR}c).

Next we simulate the effect of disoder in the leads by including 
a rather thick (400~ML) disordered region (Anderson model of disorder) 
in the left 
lead\cite{ note2} (Fig.~\ref{ fig:MR}d). This yields a significant
reduction of magneto-resistance: this is easily explained 
as the effect
of an additional resistance in series with the multilayer. The other effect 
is that the
quantum size effect is strongly perturbed. 

In order to understand better the effect of disorder on the 
quantum size 
effect, we 
show in Fig.~\ref{ fig:Gdd} the minority-spin contribution to the 
conductance of the ferromagnetic alignment
(which is the one giving rise to the quantum size effect here). We see that, 
for a 
disordered lead (Fig~\ref{ fig:Gdd}d), the amplitude of the 
conductance oscillation is reduced 
by a factor 
of 4, while the conductance itself is only reduced by 30 percent. Thus, this
suggests  that diffuse scattering in the leads supresses the quantum size
effect (here it is not completely supressed because the disordered region is
rather thin and the ballistic conductance still accounts for 70 percent of the
total conductance).

\narrowtext{
\begin{figure}[h]
%\begin{center}
\epsfxsize=8cm
\epsffile{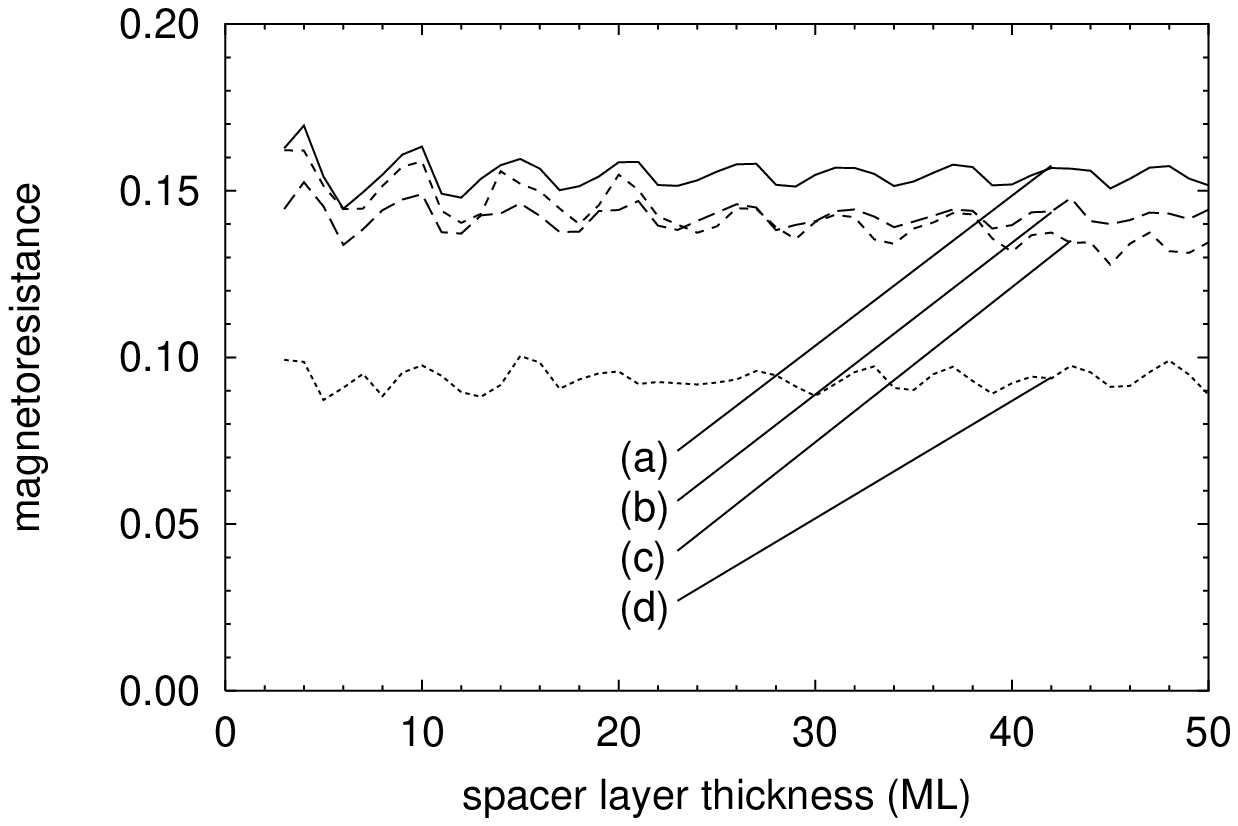}
%\end{center}
%\vspace*{\baselineskip}
\caption{Magneto-resistance. (a)
perfect case; (b) interface interdiffusion (2 layers of substitutional disorder); 
(c) bulk disorder 
in spacer layer 
(Anderson model, $W=0.5$~$|t|$); (d) 400 disordered layers in left lead (Anderson 
model,
$W=0.5$~$|t|$.} 
\label{ fig:MR}
\end{figure}
}

\narrowtext{
\begin{figure}[h]
%\begin{center}
\epsfxsize=8cm
\epsffile{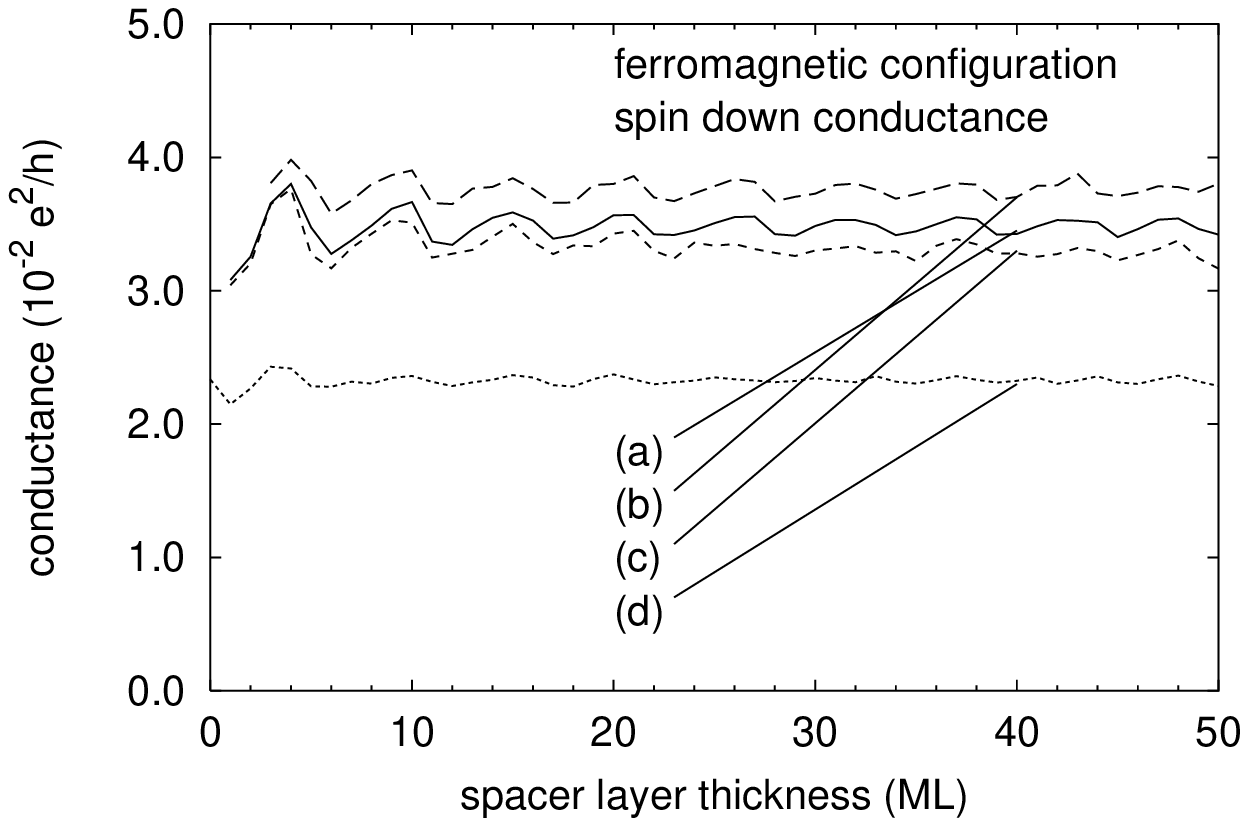}
%\end{center}
%\vspace*{\baselineskip}
\caption{Minority-spin conductance for the ferromagnetic configuration.
(a)--(d): same as Fig.~\protect\ref{ fig:MR}.}  \label{ fig:Gdd}
\end{figure}
}

\section*{Acknowledgements}
P.B. gratefully thanks the Japan Society for the Promotion of Science for 
the financial support of his stay in Nagoya during which this work was
initiated.

%\vspace*{-\baselineskip}

\end{multicols}


\begin{references}
\bibitem{ note} Note that the present definition differs from the one used by 
other 
authors, who normalize
the magneto-resistance by $G_F$ instead of $G_F+G_{AF}$ as done here; the 
present 
convention gives values strictly bound between -1 and +1, and gives numerical 
values 
that are smaller by typically a factor 2.
\bibitem{ note2} For technical reasons, disorder could not be included 
conveniently for 
both leads, 
but results with 2 disordered leads may be expected to be similar to the one 
obtained 
with 1 disordered lead.
\end{references}
\end{document}